\documentclass{blois}

\usepackage{wrapfig}

\def\be{\begin{equation}}
\def\ee{\end{equation}}
\def\bea{\begin{eqnarray}}
\def\eea{\end{eqnarray}}

\newcommand{\Photo}{}

\begin{document}
\title{More Higgses at the LHC and the Electroweak Phase Transition}

\author{G. C. Dorsch$^{1}$, S. J. Huber$^{2}$, K. Mimasu$^{2}$ and J. M. No$^{2}$}

\address{$^{1}$ DESY, Theory Group, Notkestrasse 85, D-22607 Hamburg, Germany\\
$^{2}$ Department of Physics and Astronomy, University of Sussex, Brighton BN1 9QH, United Kingdom}

\maketitle

\abstracts{
A cosmological first order electroweak phase transition could explain the origin of the cosmic matter-antimatter asymmetry. 
While it does not occur in the Standard Model, it becomes possible in the presence of a second Higgs doublet. In this context, 
we obtain the properties of the new scalars $H_0$, $A_0$ and $H^{\pm}$ leading to such a phase transition, showing
that its key LHC signature would be the decay $A_0 \rightarrow H_0 Z$, and we analyze the promising LHC search 
prospects for this decay in the $\ell \ell b\bar{b}$ and $\ell \ell W^{+} W^{-}$ final 
states. Finally, we comment on the impact of the $A_0 \rightarrow H_0 Z$ decay on current LHC searches for $A_0$ decaying into SM particles.}

\section{Introduction}

A primary goal of the Large Hadron Collider (LHC) physics programme is the study of the electroweak (EW) symmetry breaking process, its nature and properties. 
While ATLAS/CMS data from LHC Run 1 have shown that the properties of the discovered Higgs particle are compatible with 
those expected for the Standard Model (SM) Higgs boson $h$, it is possible that the EW symmetry breaking scalar sector includes more states beyond one $SU(2)_L$ doublet. 
Extensions of the SM scalar sector such as Two-Higgs-Doublet-Models (2HDMs) could explain the 
generation of the observed cosmic matter-antimatter asymmetry through EW Baryogenesis~\cite{EWBG}. A key requirement for successful baryogenesis is that the EW Phase 
Transition (EWPT) in the early Universe be strongly first order, which does not occur for the SM with a $m_h = 125$ GeV Higgs~\cite{Kajantie:1995kf}.
We show~\cite{Dorsch:2014qja} that the primary signature of a strongly first order EWPT in 2HDMs is a large mass splitting $m_{A_0} - m_{H_0}$,
leading to the decay $A_0 \rightarrow Z H_0$ as a key LHC probe of such scenario. We then discuss the upcoming LHC prospects 
in $\ell \ell\,b\bar{b}$ and $\ell \ell\,W^{+} W^{-}$ searches, and comment on the impact of the large mass splitting~\cite{Dorsch:toappear} on 
current LHC searches for $A_0/H_0$. 

Altogether, the study in~\cite{Dorsch:2014qja} highlights that the decay $A_0 \rightarrow Z H_0$, being a `smoking gun' signature of 2HDM scenarios with a strongly first order EWPT,
can be probed at the upcoming Run of LHC, thus providing a powerful connection of EW Cosmology to LHC physics.

\section{The EW Phase Transition with Two Higgs Doublets}

The 2HDM scalar sector contains two $SU(2)_L$ doublet fields $\Phi_{1,2}$ (see \cite{Branco:2011iw,Gunion:2002zf} for a review of 2HDMs). 
In addition to the recently observed Higgs boson $h$, 
the 2HDM physical spectrum then contains (in the following we assume for simplicity no Charge-Parity (CP) violation in the scalar sector) 
another neutral CP-even scalar $H_0$, a neutral CP-odd scalar $A_0$ and a charged scalar $H^{\pm}$.
After fixing the EW vacuum expectation value ({\it vev}) $v = 246$ GeV and the Higgs mass $m_h = 125$~GeV, the
remaining parameters in the scalar potential are: the physical masses $m_{H_0}$, $m_{A_0}$, $m_{H^{\pm}}$, two angles $\beta$ and $\alpha$
and a mass scale $\mu$. Here $\alpha$ is defined such that when $\alpha = \beta$, $h$ 
has SM-like couplings to gauge bosons and fermions, known as the \emph{alignment limit} (see~\cite{Dorsch:2013wja} for details on the 2HDM parameter 
definitions and conventions used in this work).

\begin{figure}[ht]
\begin{center}
\vspace{-3mm}
\includegraphics[width=0.45\textwidth, clip]{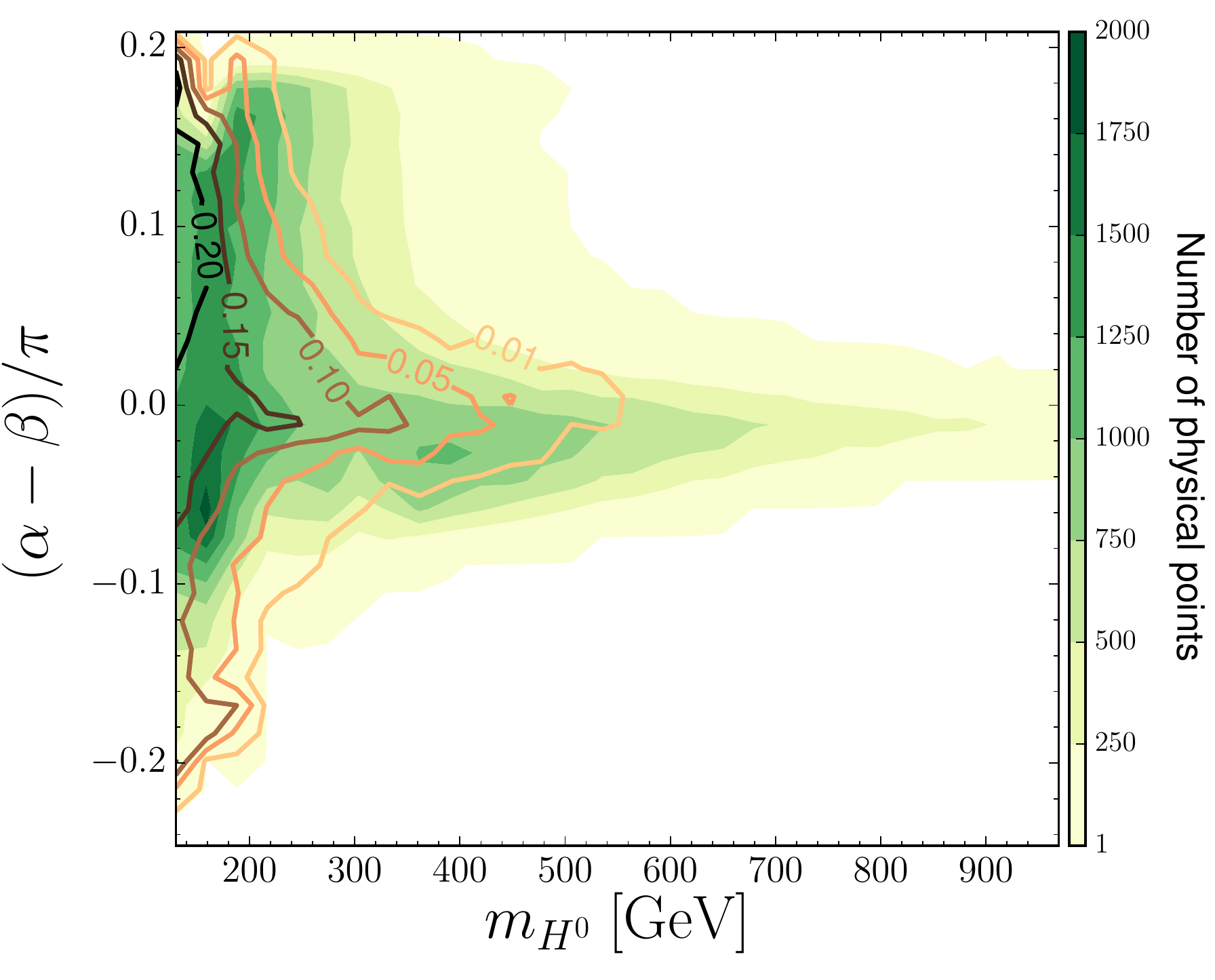}\hspace{4mm}
\includegraphics[width=0.44\textwidth, clip]{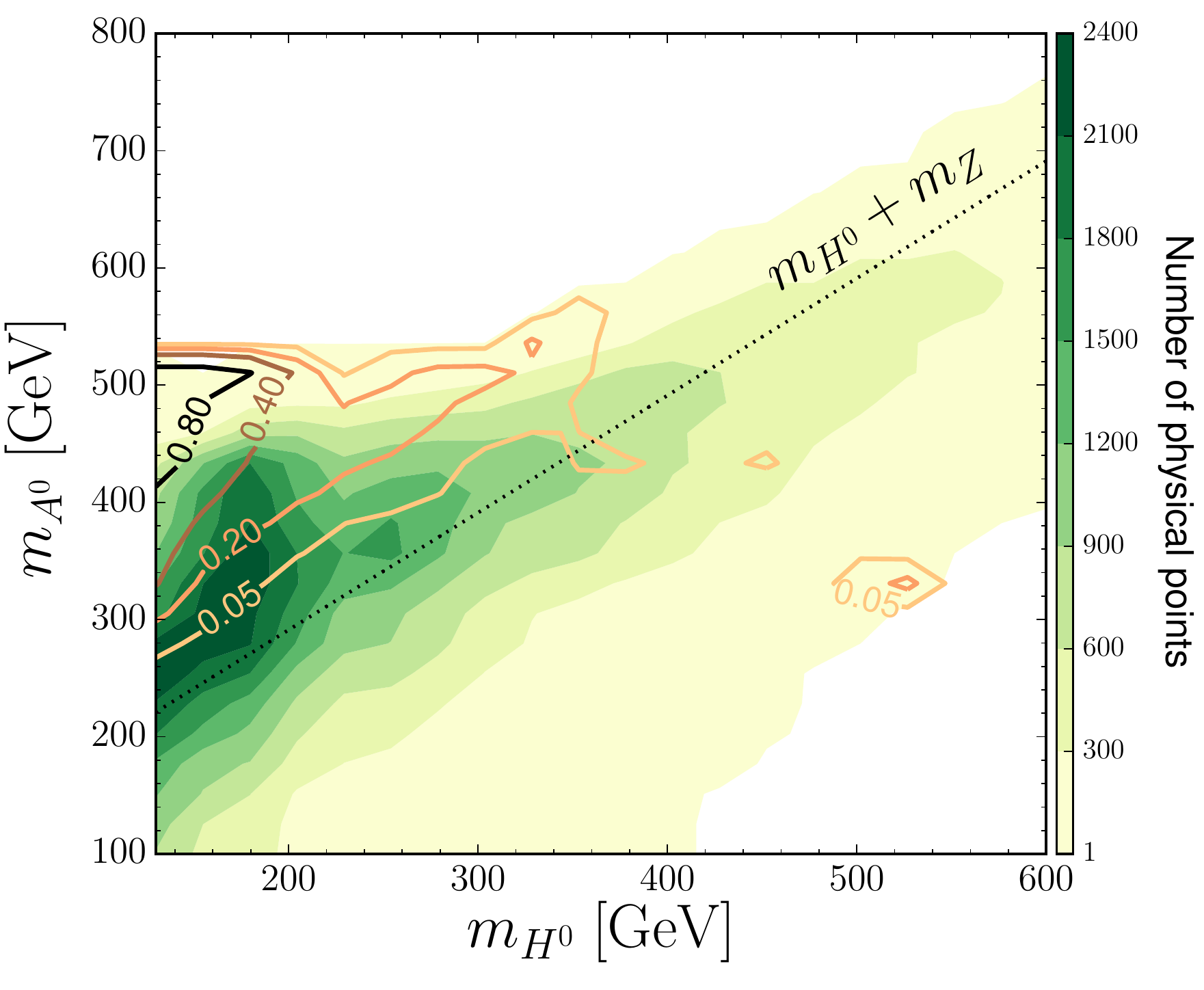}
\vspace{-5mm}
\caption{\small Heat-maps for physical points in the ($m_{H_0}, \alpha - \beta$) (\emph{left}) and ($m_{H_0}, m_{A_0}$) (\emph{right}) planes. Contours in each case show 
the region with a certain ratio of strongly first order EWPT points to physical points. The dotted-black line (\emph{right}) corresponds to $m_{A_0} = m_{H_0} + m_{Z}$.}
\vspace{-8mm}
\label{fig:EWPT}
\end{center}
\end{figure}

Our study of the strength of the EWPT in 2HDMs is performed in a Yukawa Type-I 2HDM. We stress that the Type of 2HDM considered is irrelevant for the EWPT, 
as the top quark couples to $\Phi_{1,2}$ in the same way for every 2HDM Yukawa Type scenario. However, 
experimental constraints do differ among Types~\footnote{Hence our choice of a Type-I 2HDM (instead of {\it e.g.} a Type-II 2HDM, for which up-type quarks couple to 
$\Phi_{2}$ while down-type quarks and leptons couple to $\Phi_1$) which is the least constrained one, in order to 
provide a better gauging of the impact of a first order EWPT on the 2HDM parameter space.}.
We perform a numerical scan over the parameters $m_{H_0}$, $m_{A_0}$, $m_{H^{\pm}}$, $\tan\beta$, 
$\mathrm{sin}(\alpha-\beta)$ and $\mu$, interfaced to 2HDMC~\cite{Eriksson:2009ws} and HiggsBounds~\cite{Bechtle:2013wla} to select the region of parameter 
space that satisfies EW precision constraints and existing collider bounds, as well as theoretical requirements from stability, unitarity and perturbativity. 
Flavour constraints~\cite{Mahmoudi:2009zx} and constraints 
from measured Higgs signal strengths on $\tan\beta$ and $\mathrm{sin}(\alpha-\beta)$ (see {\it e.g.}~\cite{Celis:2013rcs}) are also included.
Points in our scan satisfying all the above constraints are considered \emph{physical points}. We compute for each of them the strength of the EWPT 
via the thermal 1-loop effective potential (see~\cite{Dorsch:2013wja} for details). The results of our scan are summarized in 
Figure~\ref{fig:EWPT}, which shows heat-maps of physical points in the planes ($m_{H_0}, \alpha - \beta$) (\emph{left}) and ($m_{H_0}, m_{A_0}$) (\emph{right}),
together with contours for the ratio of strongly first order EWPT points to physical points.
A strongly first order EWPT, as needed for successful EW Baryogenesis, 
is preferentially achieved for a large mass splitting $m_{A_0} - m_{H_0} \gg m_{Z}$, together with a heavy CP-odd scalar $A_0$ ($m_{A_0} > 300$ GeV), as shown in 
Figure~\ref{fig:EWPT} (\emph{right}). Such an EWPT also favours an SM-like Higgs $h$, {\it i.e.} small $\mathrm{sin}(\alpha-\beta)$ and 
moderate $\tan\beta$~\cite{Dorsch:2014qja,Dorsch:2013wja}, with the EWPT range in $\mathrm{sin}(\alpha-\beta)$ shrinking as the 
state $H_0$ becomes heavier (see Figure~\ref{fig:EWPT} (\emph{left})). 

\section{The Decay Channel $A_0 \rightarrow Z\,H_0$ and LHC Probes of 2HDM}
\label{section3}

The large mass splitting $m_{A_0} - m_{H_0}$ leads to the $A_0 \rightarrow Z H_0$ decay channel as a characteristic LHC signature of 2HDMs with a strongly first order EWPT. 
This decay is strongly enhanced both by the large phase space available and by the coupling $g_{A_0ZH_0} \sim \mathrm{cos}(\alpha-\beta)$, 
unsuppressed in the alignment limit (in contrast with the decay $A_0 \rightarrow Z h$, which vanishes in that limit since $g_{A_0Zh}
\sim \mathrm{sin}(\alpha-\beta)$). Regarding competing decay channels, $A_0 \rightarrow t\bar{t}$ is 
subdominant for $m_{A_0}-m_{H_0} > v$, while the presence of $A_0 \rightarrow W^{\pm} H^{\mp}$ depends on the splitting $m_{A_0}-m_{H^{\pm}}$. 
EW precision observables require $H^{\pm}$ to be close in mass to either $H_0$ or $A_0$ \cite{Grimus:2007if},
which makes $A_0 \rightarrow W^{\pm} H^{\mp}$ either kinematically forbidden or similar to $A_0 \rightarrow Z H_0$ (see~\cite{Dorsch:2014qja} for a more detailed discussion 
on these issues). Here we assume for simplicity $m_{H^{\pm}} \sim m_{A_0}$. 

In the following we analyze two prototypical scenarios, featuring $\mu = 100$~GeV, $\mathrm{tan}\,\beta = 2$, $m_{A_0} = m_{H^{\pm}} = 400$~GeV, $m_{H_0} = 180$~GeV and respectively 
$(\alpha - \beta) = 0.001\,\pi$ (Benchmark A) and $(\alpha - \beta) = 0.1\,\pi$ (Benchmark B).
These benchmarks characterize the two alternatives for the dominant decay of $H_0$: $H_0 \rightarrow b \bar{b}$ very close to the alignment limit, while 
away from it $H_0\rightarrow W^+W^-$.
This discussion highlights the fact that for 2HDMs with a strongly first order EWPT, the corresponding ``smoking gun" signature at the LHC will 
either be $p p \rightarrow A_0 \rightarrow Z H_0 \rightarrow \ell \ell b \bar{b}$ or $p p \rightarrow A_0 \rightarrow Z H_0 \rightarrow \ell \ell W^+W^-$, depending 
on how close the 2HDM is to the alignment limit. 

\subsection{LHC Search for $A_0$ in $\ell\ell\,b\bar{b}$ and $\ell\ell\,W^{+}W^{-}$}

\begin{figure}[t]
\begin{center}
	\includegraphics[width=0.48\textwidth, clip]{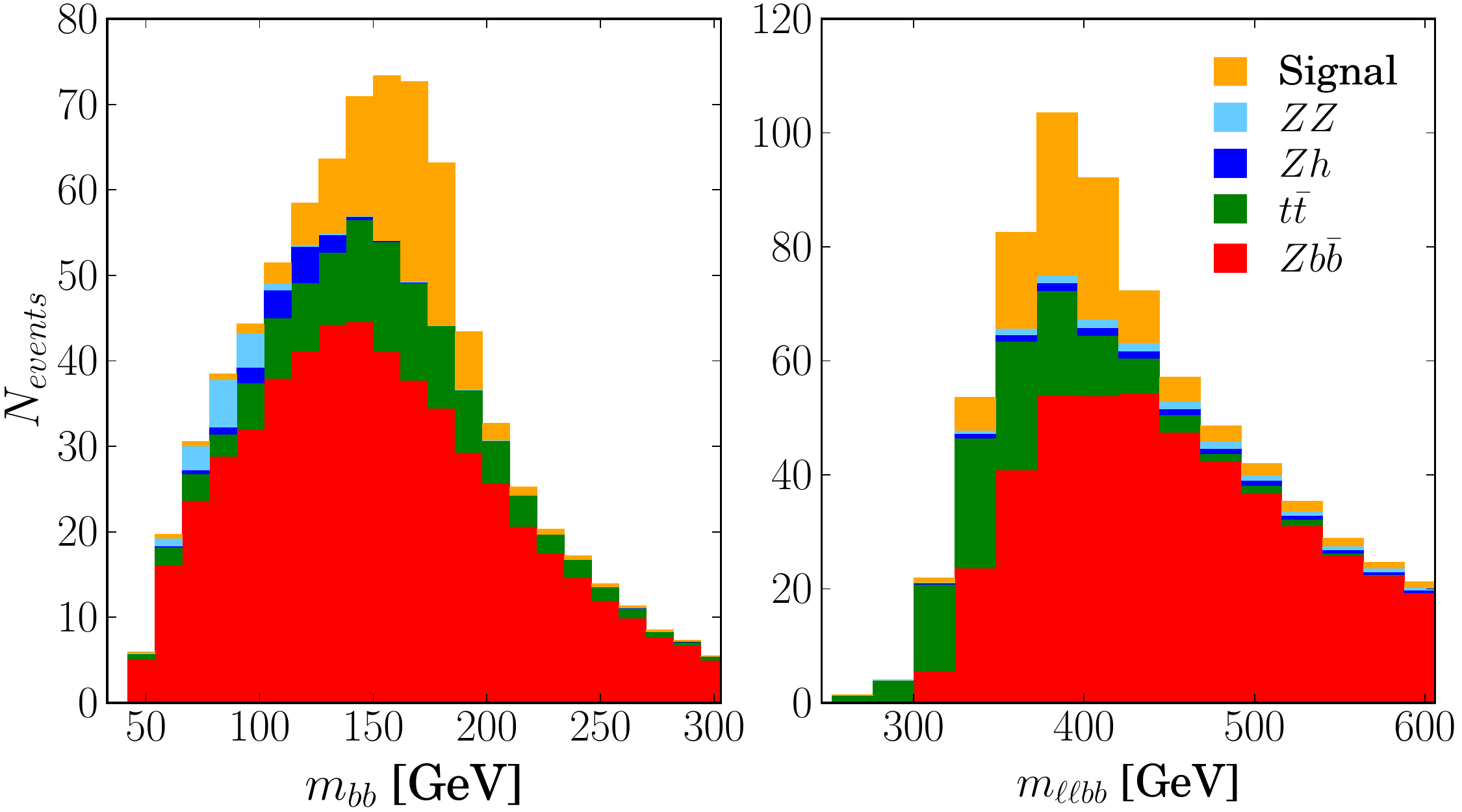}\hspace{3mm}
	\includegraphics[width=0.48\textwidth, clip]{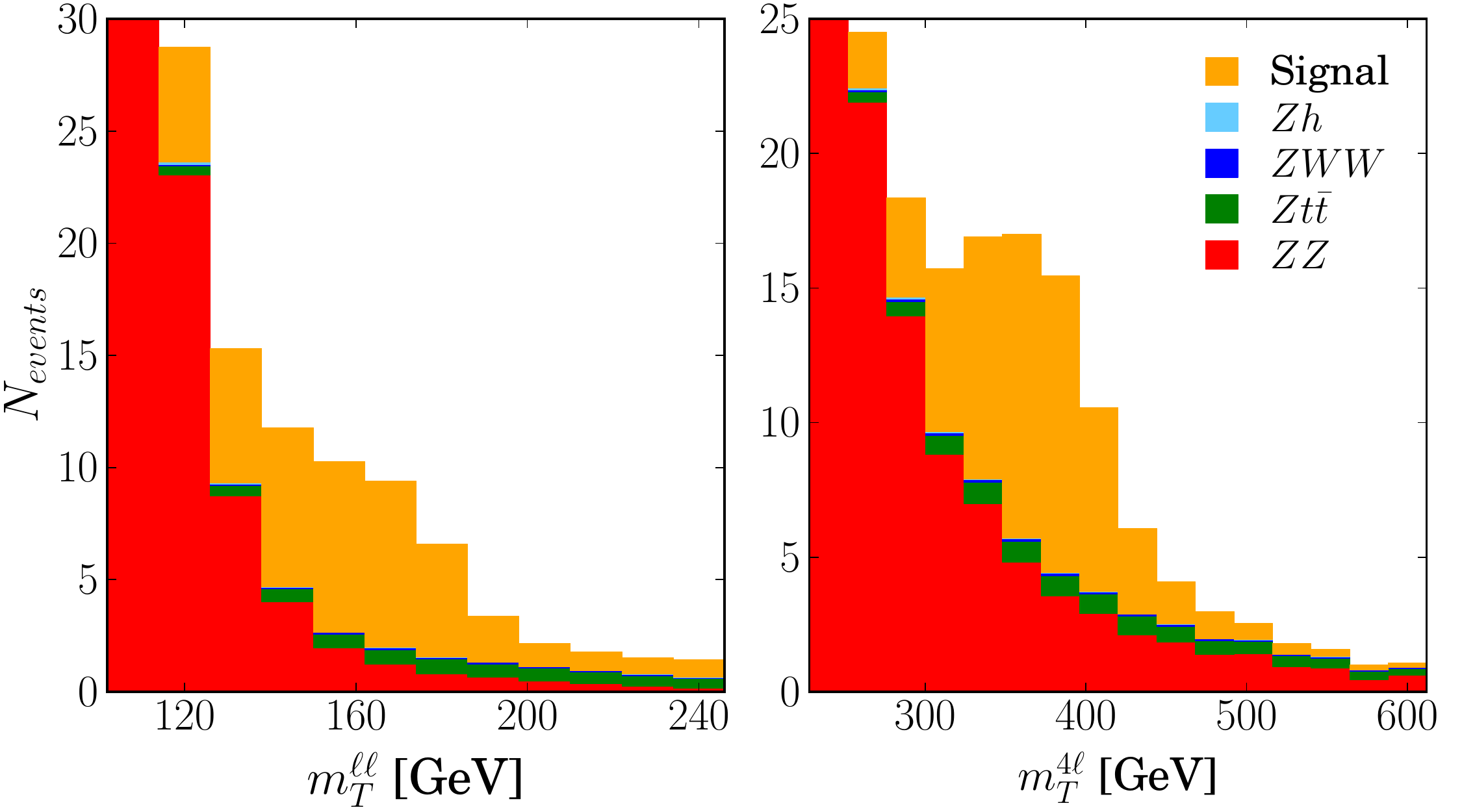}
	\vspace{-3mm}
	\caption{\small Left: $m_{bb}$ (\emph{left}) and $m_{\ell\ell bb}$ (\emph{right}) distributions after analysis cuts, 
	with the various contributions stacked (for $\mathcal{L} = 20\,\, \mathrm{fb}^{-1}$). Right: 
	$m^{\ell\ell}_{T}$ (\emph{left}) and $m^{4\ell}_{T}$ (\emph{right}) distributions after event selection, with the 
	various contributions stacked (for $\mathcal{L} = 60\, \,\mathrm{fb}^{-1}$).}
	\vspace{-5mm}
	\label{fig:dist1}
	\end{center}
\end{figure}

We now analyze the search prospects in the $\ell \ell b \bar{b}$ and $\ell \ell W^+W^-$ channels at the 14 TeV run of the LHC using the defined benchmarks A and B from 
above. We implement the Type-I 2HDM in {\sc FeynRules}~\cite{Christensen:2008py} 
and use {\sc MadGraph5$\_$aMC$@$NLO}~\cite{Alwall:2011uj} to generate both signal and background analysis samples, then passed on to {\sc Pythia}~\cite{Sjostrand:2007gs} 
and {\sc Delphes}~\cite{deFavereau:2013fsa} for parton showering, hadronization and detector simulation.
We first concentrate on Benchmark A, which corresponds to the $\ell \ell b \bar{b}$ final state. The two main SM backgrounds are: (i) $Z b\bar{b}$ production 
(with $Z \rightarrow \ell \ell$), (ii) QCD $t\bar{t}$ production (with $t\bar{t} \rightarrow b W^{+} \bar{b} W^{-} \rightarrow b \ell^{+} \nu_{\ell} \bar{b} \ell^{-} \bar{\nu}_{\ell}$).
Event selection requires the presence of two isolated same flavour (SF) leptons in the final state with $P^{\ell_1}_{T} > 40$ GeV, $P^{\ell_2}_{T} > 20$ GeV and $\left| \eta_{\ell} \right| < $ 2.5 
(2.7) for electrons (muons), together with two b-tagged jets in the event with $P^{b_1}_{T} > 40$ GeV, $P^{b_2}_{T} > 20$ GeV and $\left| \eta_{b} \right| < $ 2.5.
In order to extract the signal we require $m_{\ell\ell} =  m_Z \pm 10$ GeV and perform the cuts (see \cite{Dorsch:2014qja} for details) 
$H_T^\mathrm{bb} > 150$ GeV, $H_T^\mathrm{\ell\ell bb} > 280$ GeV, $\Delta R_{bb} < 2.5$,
$\Delta R_{\ell\ell} < 1.6$. 
We define the signal region as $m_{bb} = (m_{H_0} - 20) \pm 30$ GeV and $m_{\ell\ell bb} = (m_{A_0} - 20) \pm 40$ GeV 
and show the $m_{bb}$ and $m_{\ell\ell bb}$ distributions after cuts for an integrated luminosity $\mathcal{L} = 20\, \mathrm{fb}^{-1}$ in Figure \ref{fig:dist1} (\emph{left})
(various contributions stacked). A discovery value $S/\sqrt{S+B} = 5$ ($S =$ signal events, $B =$ background events) may be obtained 
already with $\mathcal{L}\sim 15-20$ fb$^{-1}$ in the limit that only statistical uncertainties are important.
\begin{figure}[h]
\begin{center}
\includegraphics[width=0.95\textwidth, clip]{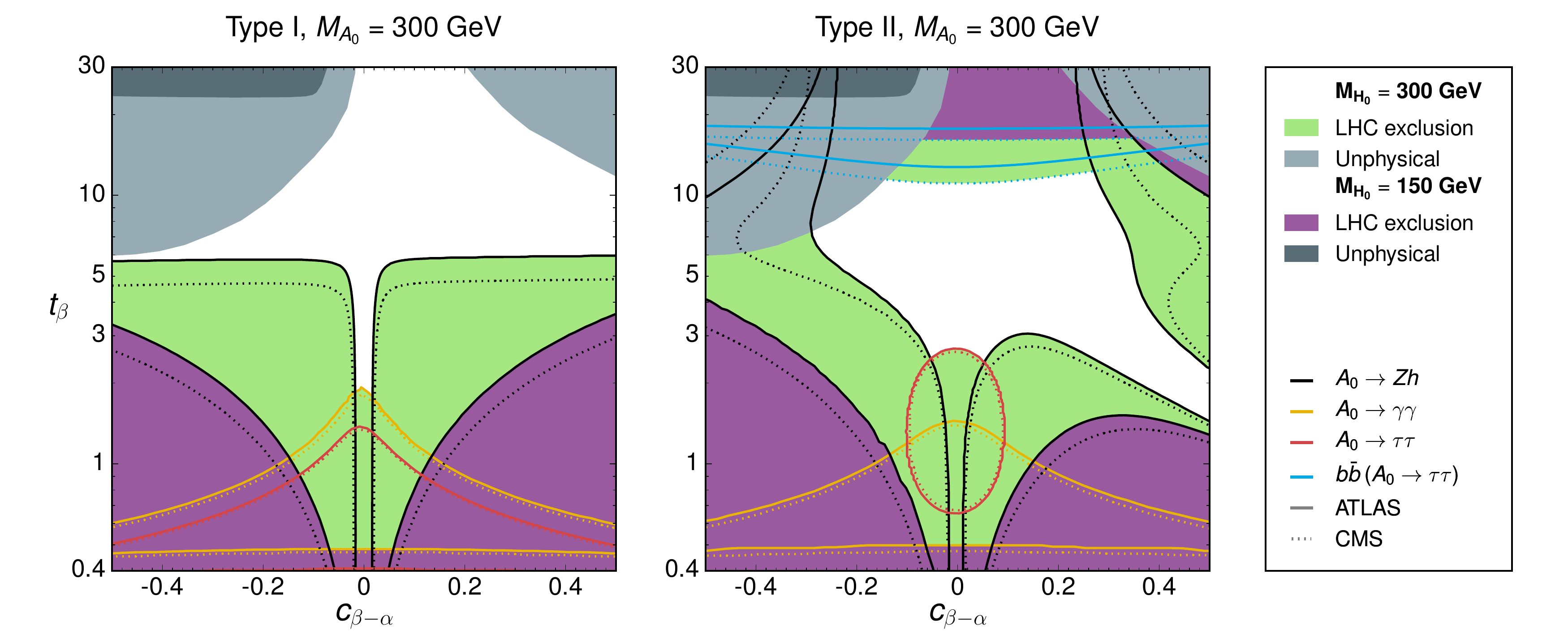}
\vspace{-4mm}
\caption{\small Current limits in the ($\mathrm{sin}(\alpha-\beta),\, \mathrm{tan}\,\beta$) plane from ATLAS/CMS searches of $A_0$ decaying into 
SM states for 2HDM Type-I (\emph{left}) and Type-II (\emph{right}), with $m_{A_0} = 300$ GeV and $m_{H_0} = 300$ GeV (light-green exclusion), $150$ GeV (purple exclusion) 
respectively.}
\vspace{-3mm}
\label{fig:limits}
\end{center}
\end{figure}

Away from alignment (Benchmark B) the decay $H_0 \rightarrow W^+W^- \to \ell \nu_{\ell} \ell \nu_{\ell}$ (together with $Z \to \ell' \ell'$)  
provides the best discovery prospects. The main background is $Z Z$ production with $Z Z \rightarrow \ell \ell \ell' \ell'$. For 
event selection, we require four isolated leptons in the final state with $P^{\ell_1}_{T} > 40$ GeV, $P^{\ell_2,\ell_3,\ell_4}_{T} > 20$ GeV, with
one SF lepton pair (opposite sign) reconstructing $m_Z$ within $20$ GeV. After event selection, the signal and background cross sections at LHC 14 TeV (at leading order)
are respectively $0.93$ fb and $5.6$ fb. Defining the transverse mass variables $m^{\ell\ell}_{T}$ and $m^{4\ell}_{T}$
\begin{equation}
\left(m^{\ell\ell}_{T}\right)^2 =  \left(\sqrt{p^2_{T,\ell\ell}+m^2_{\ell\ell}}  + \slash\hspace{-2mm} p_{T}\right)^2 - \left(\vec{p}_{T,\ell\ell} + 
\slash\hspace{-2mm}\vec{p}_{T}\right)^2 \, \, ,\quad 
m^{4\ell}_{T} = \sqrt{p^2_{T,\ell'\ell'}+m^2_{\ell'\ell'}} + \sqrt{p^2_{T,\ell\ell}+\left(m^{\ell\ell}_{T}\right)^2}
\end{equation}
with $\ell'\ell'$ the two SF leptons most closely reconstructing $m_Z$, a signal region of $m^{4\ell}_{T} > 260$ GeV 
(see Figure \ref{fig:dist1} (\emph{right})) allows to extract a clean signal \cite{Dorsch:2014qja}. Our final signal cross section is 1.41 fb, which compared 
to a background of 1.7 fb reaches a significance of 5 with $\mathcal{L}\sim 60$ fb$^{-1}$.

\subsection{Impact of $m_{A_0}-m_{H_0}$ on LHC Searches for $A_0$ into SM States}

Finally, we stress that the presence of the dominant decay mode $A_0 \rightarrow Z H_0$ due to a large splitting $m_{A_0}-m_{H_0}$ 
has an important impact on the sensitivity of current searches for $A_0$ decaying 
into SM states at the LHC, as the branching fractions into those get significantly reduced for a large mass splitting~\cite{Dorsch:toappear} $m_{A_0}-m_{H_0}$. 
As an example, Figure \ref{fig:limits} shows the current limits in the 2HDM parameter plane ($\mathrm{sin}(\alpha-\beta),\, \mathrm{tan}\,\beta$) from ATLAS/CMS searches of 
$A_0$ decaying into SM states~\cite{Aad:2015wra}, for $m_{A_0} = 300$ GeV and $m_{H_0} = 300$ GeV, $150$ GeV respectively, for 2HDM Type-I (\emph{left}) and Type-II (\emph{right}).

\vspace{-3mm}
\section*{Acknowledgments}
\vspace{-4mm}

S.H. and K.M. are supported by the Science Technology and Facilities Council (STFC) under grant number ST/L000504/1.
The work of J.M.N. is supported by the People Programme (Marie curie Actions) of the European Union Seventh Framework Programme (FP7/2007-2013) 
under REA grant agreement PIEF-GA-2013-625809.

\end{document}